# Production Line Technique for Autonomous Vehicle Scheduling

Nasser Aloufi
Department of Computer Science
California State University, Dominguez Hills
Carson, CA USA
naloufi1@toromail.csudh.edu

**Abstract - This paper considers the problem of scheduling autonomous vehicles in intersections. A new system is proposed that could be an additional choice to the recently introduced autonomous intersection management (AIM) model. The proposed system is based on the production line technique, where the environment of the intersection, vehicles position, speeds and turning are specified and determined in advance. The goal of the proposed system is to eliminate vehicle collision and the waiting time inside the intersection. 3 different patterns of the vehicles' flow toward the intersection have been considered for the evaluation of the model. The system requires less waiting time –compared to the other models- in the random case where the flow is unpredictable. The KNN algorithm is used to predict the right turn vehicle. The experimental results show that there is no single chance of collision inside the intersection; however, the system requires more free space in the traffic lane.**

## I. Introduction

The number of vehicles on streets is increasing day after another. The more vehicles we have in roads, the more furious people get which resulting in breaking traffic rules and delaying the vehicles behind. As vehicles are expecting to be completely autonomous, a new solution to traffic jams has sparked. The necessity of creating an autonomous vehicles and intelligent transportation systems is more than ever before. This intelligent system can be done by making vehicles interact with each others and updating their routes according to the traffic flow.

If we look at the history of autonomous vehicle we can see that they went through 5 stages [1]: A) No Automation: in this stage the human was responsible about all the driving. No automation interaction was involved. B) Driver Assistance: in this stage the system was responsible about simple tasks such as steering and acceleration/deceleration. C) Partial Automation: this stage is a continuation of the previous one with some enhancement. Features such as cruise control and lane-centering have been added to the vehicles. D) Conditional Automation: the autonomous driving system performs all aspect of tasks; however, the human should be standing by, in case anything went wrong. E) High Automation: in this stage, the vehicle operates autonomously to all conditions in the domain. Any scenario in the system domain, the vehicle should do it fully autonomously. F) Fully Automation: this is the final stage where human intervention is not needed. The system's performance is equal to human driver in all scenarios without any domain constraints. It is expected for all vehicles to reach the fully automation stage and work without any human control by the year of 2035 [2]. The Institute of Electrical and Electronics Engineers (IEEE) have claimed that driverless vehicles will be the most viable form of intelligent transportation. They estimate that up to 75% of all vehicles around the world will be autonomous by the year of 2040 [3].

In 2008, The Defense Advanced Research Projects Agency (DARPA) held an event which ask teams to build autonomous vehicles that have the ability to drive in traffic, make maneuvers and park. The event was described as the first time that autonomous vehicles make interaction between manned and unmanned vehicles in a real environment [4]. The VisLab in Italy that had made many advanced driver assistance systems [5], [6] [7] [8], and prototype vehicles including ARGO [9], TerraMax [5], [10], [11], and BRAiVE [12], [13] made another experiment related to the autonomous vehicle development. The team that consists of Massimo Bertozzi, Alberto Broggi, Alessandro Coati, and Rean Isabella Fedriga made a long trip experiment for four autonomous vehicles [14]. The trip took a place from Italy to China crossing more than 15,000 km for a consecutive three months. The result of their experiment told us three major things: a- the autonomous vehicle faced troubles while making maneuver; b- being one of the few vehicles that follow the street rules might cause a long waiting time (specially while applying FIFO); c- hard to combine autonomous with un-autonomous vehicles in the same environment. In [15], an intersection algorithm model was proposed. The model is based on (MixedInteger Linear Program) MILP controller. Their contributions are: 1) an algorithm that predict vehicle arrivals. 2) Applying mixed-integer linear program. 3) Develop a simulation that based on the proposed MILP controller. For a maneuver and changing lanes, a politely change lane (PCL) [16] paper was proposed. The main goal of the PCL is to provide safety and efficiency while maneuver.

In 2013, Mladenović and Abbas proposed a self-organizing control framework for driverless vehicles [27]. Their proposal is based on cooperative control framework and intersections as agent system, where distributed vehicle intelligence has been used to get the vehicle's velocity. The priority level system determines the vehicle that pass the intersection first and adjust the speed for the following vehicle in the queue so they don't collide. In 2014, Yan, Wu, and Dridi [28] studied the complexity of the sequencing problem. Their simulation was based the AIM framework. They converted the autonomous vehicles scheduling to a single machine scheduling problem, then they proved that it is an NP-hard problem. Dresner and Stone proposed a new intersection control mechanism called Autonomous Intersection Management (AIM) [29]. Their paper shows that by making a smart intersection controlling system, vehicles' flow would be more efficient than the current situation (traffic signals and stop signs). They suggested that the drivers and the designed intersections should be treated as agents. The system could have more than one agent leading us to have a Multiagent Systems (MAS). The MAS includes all the interacted elements in the environment such as drivers, pedestrians, speeds and road signs. Whenever a vehicle wants to reserve a place in the intersection, it sends a request to the intersection manager, and then the intersection manager takes an action by weather accepting or rejecting the vehicle's request. Figure1 below shows how the AIM system works[29].

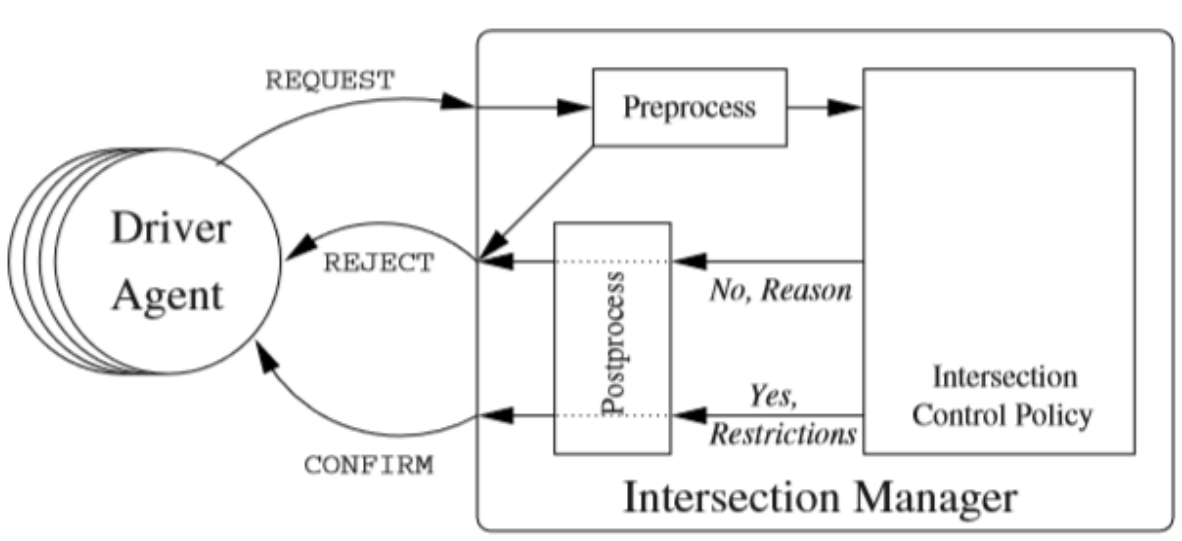

Figure1.How an AIM driver agent make a reservation.

In this paper, we propose a new scheduling model that based on the producion line technique where items don't collide and interfere with each other. Our proposed method can be implemented for faster computation [19][20] using GPUs as well [21][25][26]. Also, the network data overhead can be reduced using compression techniques [18][22][23], or using cloud computing resources [24].

The rest of the paper is organized as the following: In section 2 we made a simulation to show the cahnce of collision and the amount of waiting time in the current model. In section 3 we propose our model. Then we provide the simulation and the result of our system.

## II. Problem Description

It seems a mission impossible to make all vehicles go through an intersection safely. We need to take into account the speed, timing, distance, the chance of collision, and the appropriate response that we need to make if there is a chance of collision. In AIM and similar systems, the vehicle sends a request to the intersection manager asking for a permission to go through the intersection. The intersection manager must make one of those decisions: 1) Accept vehicle's request (when the vehicle meets the requirements). 2) Reject vehicle's request (when the vehicle failed to meet all the requirement). The rejected request has two options too: 1- ask the vehicle to accelerate or decelerate its speed (in this case the vehicle should send a new request to the intersection manager. 2- ask the vehicle to stop due to requirements failure. Figure2 below shows a general overview of the current intersection scheduling model.

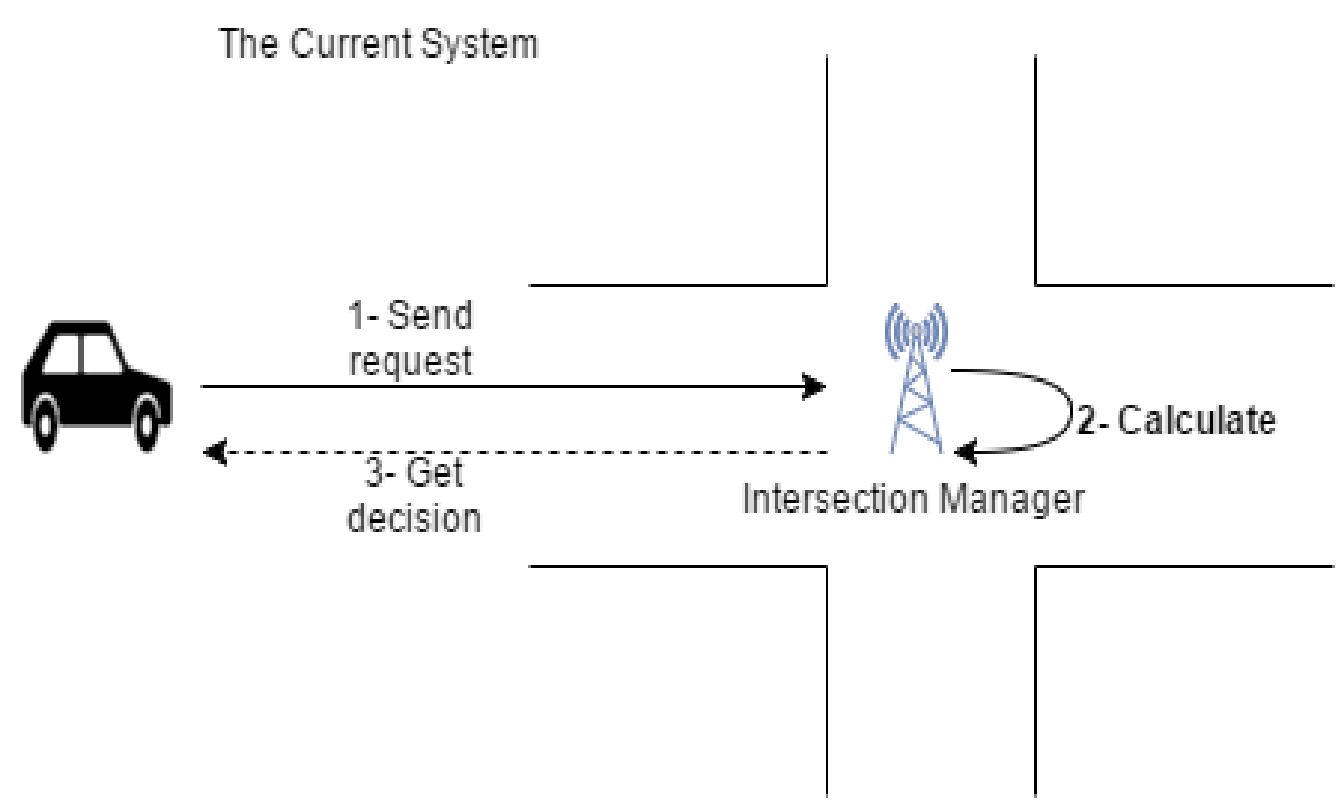

Figure2. The current system structure.

The current model works well, but still has some problems that we have to overcome before applying it to a real autonomous environment. Asking vehicles to send requests few hundred meters before approaching the intersection cause a critical time situation. There are some condition where stopping the vehicle completely is a must, and this happens whenever:

$V_x(T,S,P) = V_z(T,S,P)$

*Where: T is the time of approaching a specific point in the intersection ,S is the speed of the vehicle,P is the occupied point in the intersection.*

The more the intersection manager asks vehicles to stop, the more traffic jam happens which make the system inefficient especially in big cities. Figure3 shows one of the conditions that cause traffic jam. Let's make an example and assume that we have two vehicles. Vehicle A is going to the east side with a speed of 80mph, and vehicle B going to the west side with the same speed. Both have the same distance from the intersection, 600 ft. Both had sent requests to the intersection manager asking to go through the intersection. The intersection manager calculated the characteristics and the conditions of the two vehicles and came to conclusion that both vehicles will approach the same point at the same time, specifically after 5.11 seconds (converting 80 mph to foot -equal to 422400- then divided by 60 –equal to 7040- then divided by 60 –equal to 117.33- foot per second). To avoid the collision, the intersection manager has two options as we stated before; either reject the requests and ask vehicle A or B to change its speed; or simply ask one of them to stop in order to avoid the collision.

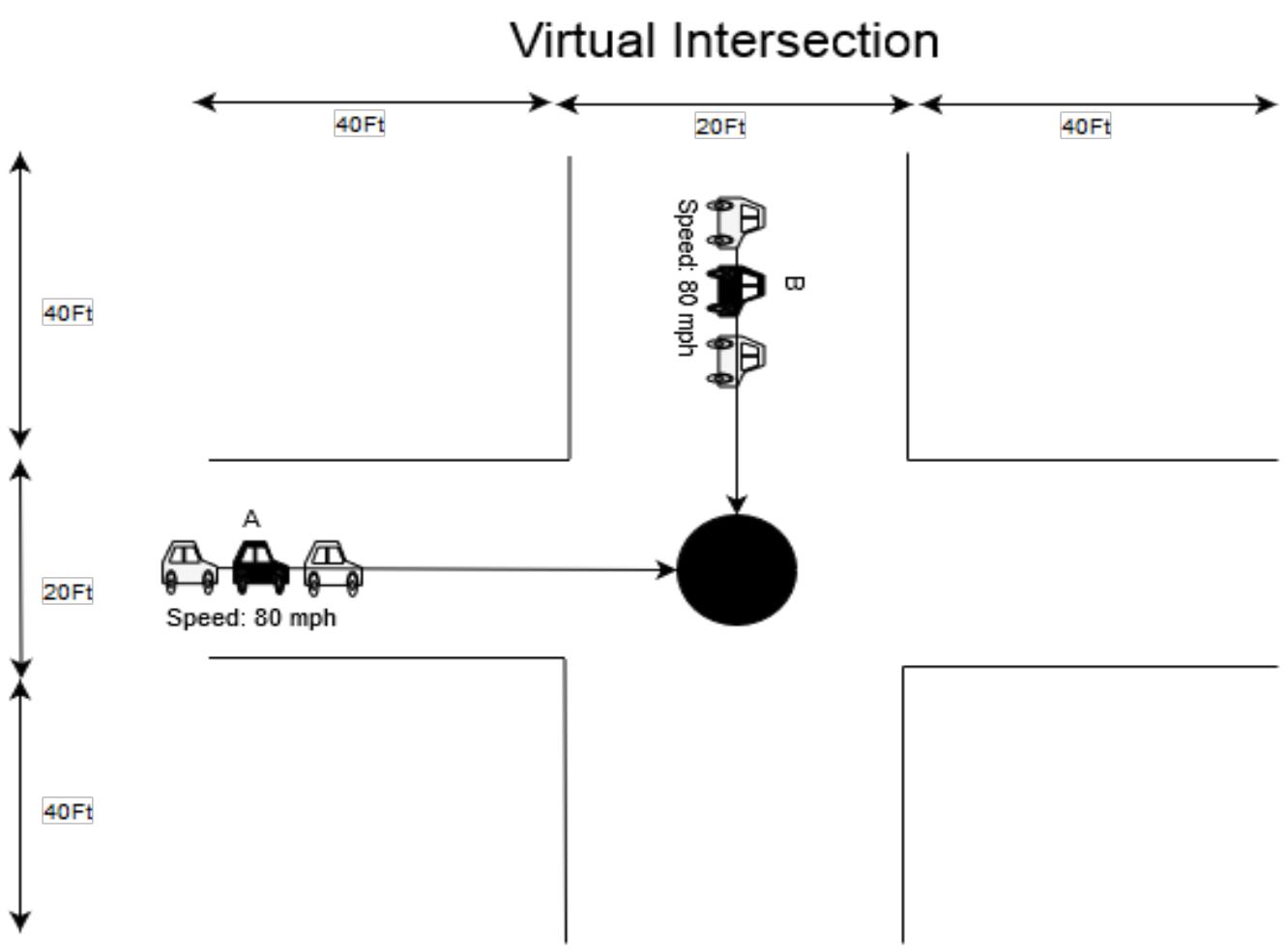

Figure3. A point of collision between vehicles A and B.

The arisen problem is, during rush hours when the intersection gets hundreds or thousands of requests, the chances of collisions increase, leading to more stopping and waiting times.

We made a simulation of a similar environment to this model in order to show the chance of collisions that each vehicle might have. The simulated intersection has the following characteristics:

I. Two directions (one direction is going from north to south and the other one is going from west to east.).
II. Each side of the intersection can occupy 722 vehicle in total.
III. All vehicles have the same speed 100 mph.

After running the simulation for 100 times, the experiment shows that, as the number of vehicles increase, the number of expected collisions and waiting times increase. When we had 50 vehicles in the intersection, the number of expected collisions was 1 collision for each vehicle. Then it was 3 when we got 200 vehicles. When we had 300 vehicles, the number of collisions increased to be 4.3 for each vehicle. The same thing happened for the waiting time. It was increasing as the number of vehicles increase. The waiting time was 85 seconds per vehicle when we had 50 vehicles inside the intersection. Then it increased to be 320 seconds when we had 200. In the end when we had 300 vehicles, the waiting time was 515 seconds per vehicle. Figure4 and Figure5 show the number of collisions and waiting times respectively.

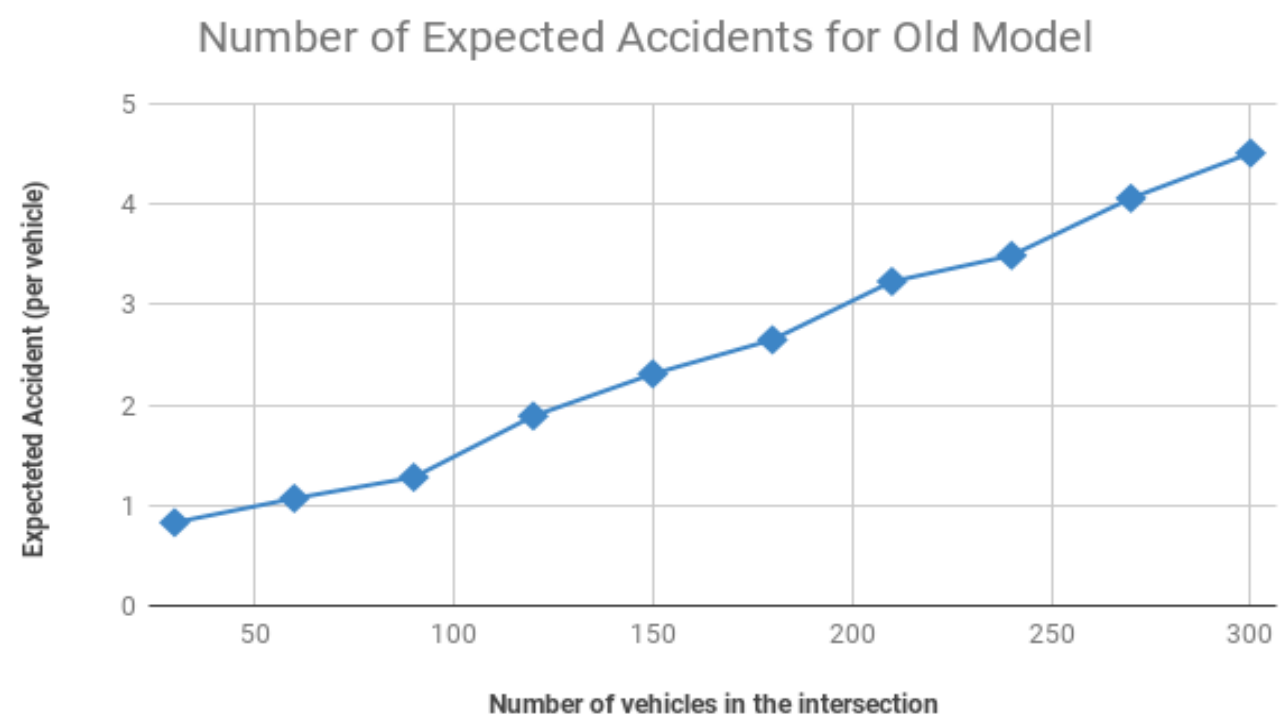

Figure4. Chances of collisions for every vehicle.

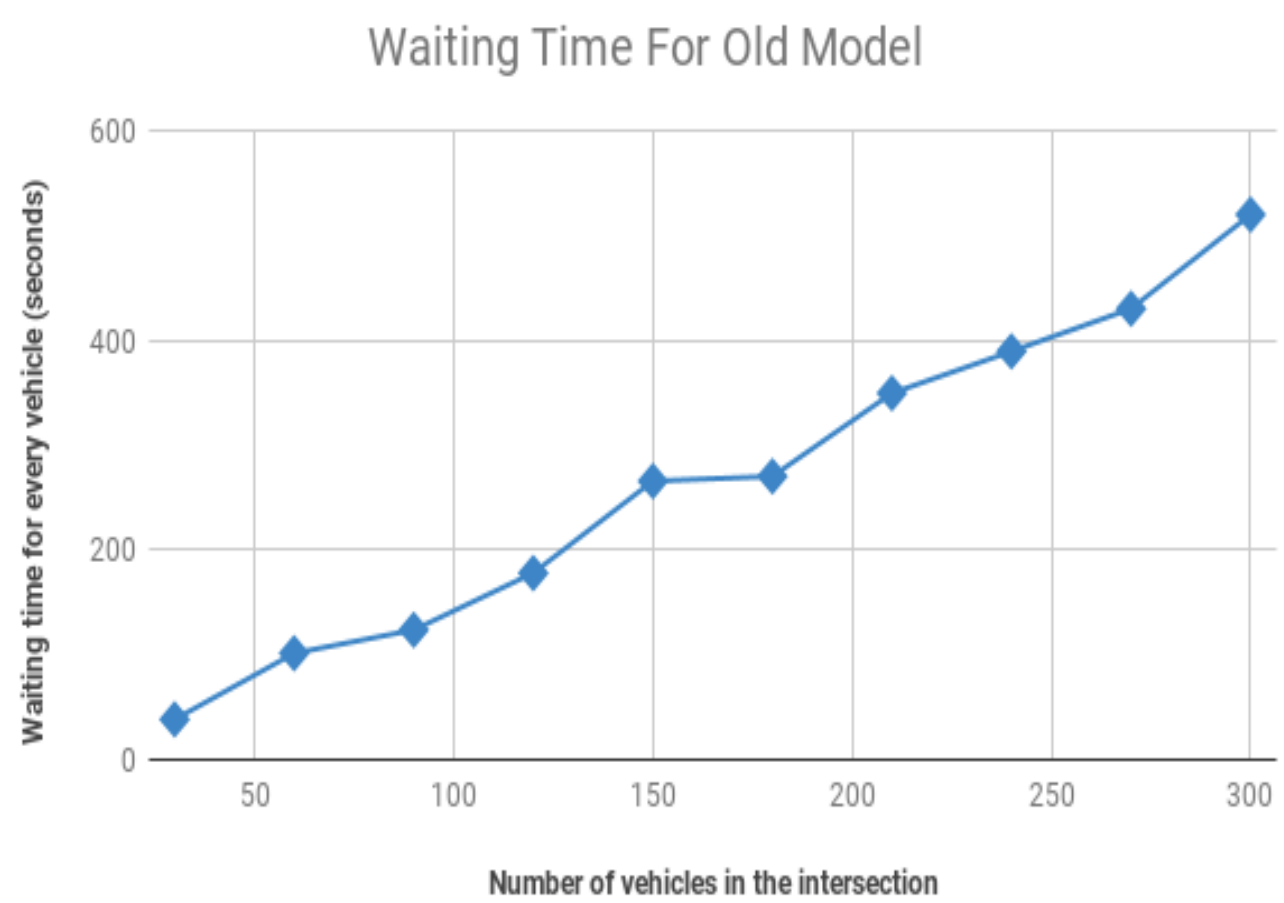

Figure5. The waiting time for every vehicle.

Increasing and decreasing speed is not an option when the intersection get enormous requests because there are vehicles in front and behind each vehicle, and whenever we fix one collision, another one appears. To sum up, the more vehicles approaching the intersection we have, the more chances of collision we get, leading us to more stopping time.

## III. The Proposed System

In the old model, the scheduling process starts after receiving the vehicle's request (Figure2). In this paper we flipped the current approach and created another model where the scheduling should be sat up in the intersection prior receiving any request. Our proposal is based on the production line system where every position (container) in the line is reserved for a specific item.

Figure 6 shows the architecture of the proposed system where the intersection's spots should be sat up as the first step. We prepare the intersection by making containers that based on the length of the vehicles. Once setting up the vehicles position, entering timing, and the speed of the lanes; then vehicles can send requests to the intersection manager.

Figure7 shows a closer look of the design where:

- S1; the minimum accepted speed.
- S2; the maximum accepted speed.
- Spin; the times where the lane is open.

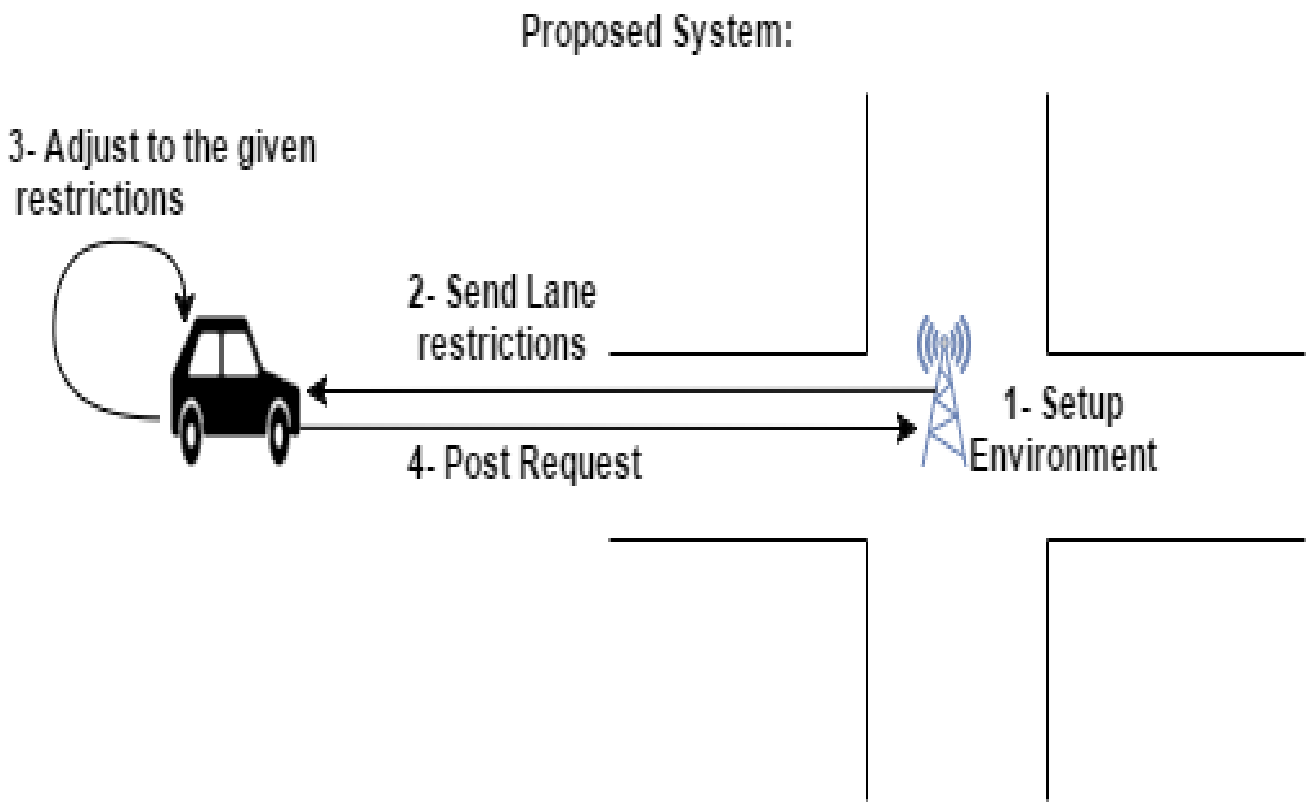

Figure6: The architecture of the proposed system.

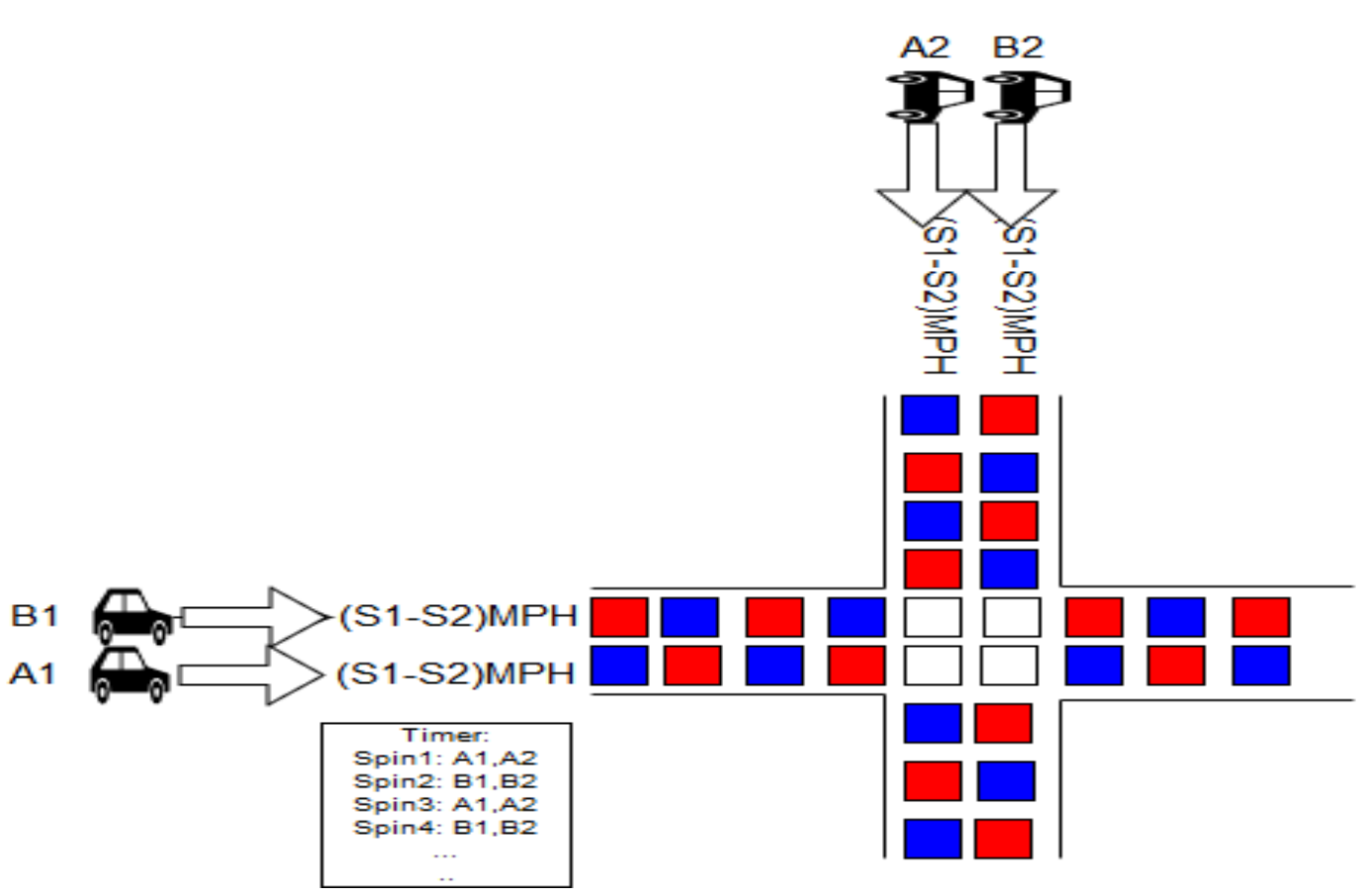

Figure7. Production line intersection.

The timer is used to switch between lanes. If lane A is open at this moment, lane B should be closed. In [30] a setpoint system for generating setpoints for the PID controllers was proposed. The purpose of the system is to make sure that the vehicle arrives at the exact expected time with the exact expected speed. However, the system doesn't guarantee the arrival time of the vehicle to a specific point, so it's not ideal where an unexpected latency cause a catastrophic result. We took that into consideration, and instead of applying one specific speed value, we applied S1 and S2 which are the minimum and maximum speed. Any vehicle speed that meets this range should be accepted.

## A) Arisen problems:

There are 2 problems that had arisen after the initial design:

1. *After applying S1 and S2, vehicles speed are vary, leading vehicles to collide after certain distance.*

Let's assume we sat up the entering speed for lane A to 60MPH as a minimum speed (S1), and to 65MPH as a maximum speed (S2). Two vehicles (V1 and V2) sent requests to enter the intersection. V1 came first with a 61MPH speed, and V2 came after it with a 64MPH speed. The problem is after a certain distance, V2 will approach V1 and collide with it.

2. *Entering an intersection with one speed range while the following intersection requires a different speed range.*

Let's assume we have two intersections, one after another. The first intersection accept vehicles with any speed between 60 and 65, and the second one accept vehicles with 102.5 - 107.5 speed range. The problem now is how to make those two intersections compatible so they get connected together in one system. Figure9. Shows this problem.

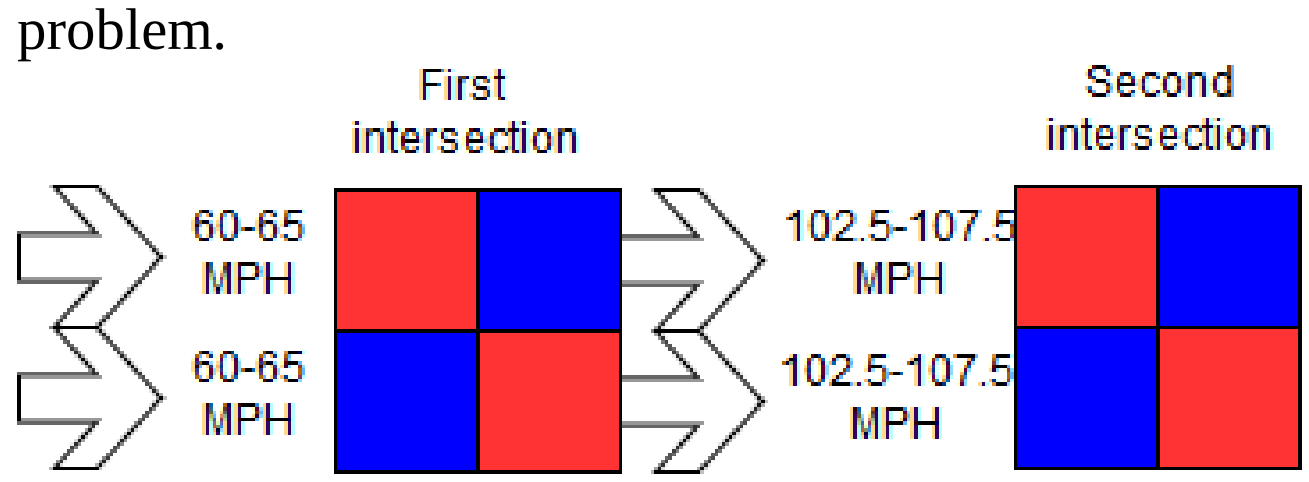

Figure9. Varying speed between two intersections.

3. *Making a right turn.*

The system must produce a container that makes a right turn, otherwise vehicles will be forced to go in straight direction.

## B) Problems solving

To solve the first problem we applied an average speed value. This average speed should be applied to every vehicle as soon as entering the intersection.

$$Avg = (S1+S2)/2$$

The goal of the average speed it to make sure that the vehicle stays in its given container and doesn't jump to another one. Figure10 shows the problem before and after applying the average speed.

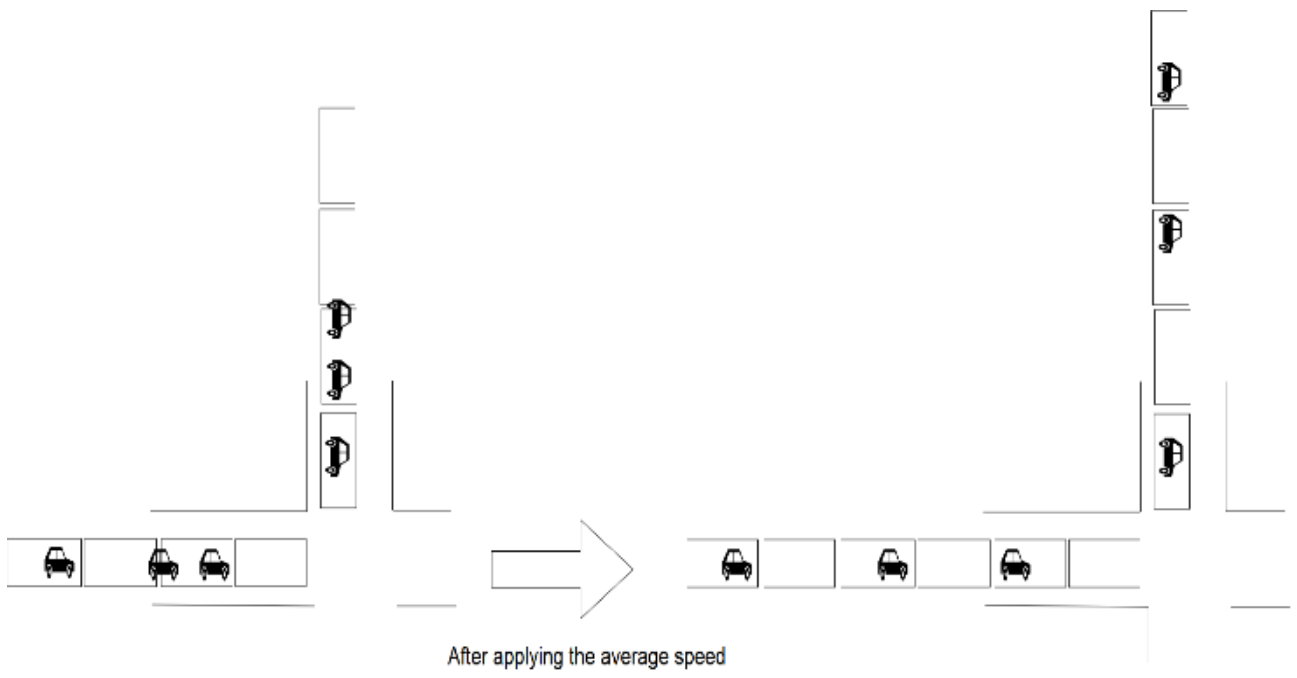

Figure10. The intersection flow before and after applying the average speed.

To solve the second problem, we created an increasing and decreasing speed areas. The intersection manager ask all the vehicles either to increase or decrease the speed depending on the speed of the next intersection.

S = Ep1 + (Tp1- Ep1); where: Ep is the exit speed of the intersection. Tp is the entering speed of the following intersection. Figure11 shows how the increasing speed works.

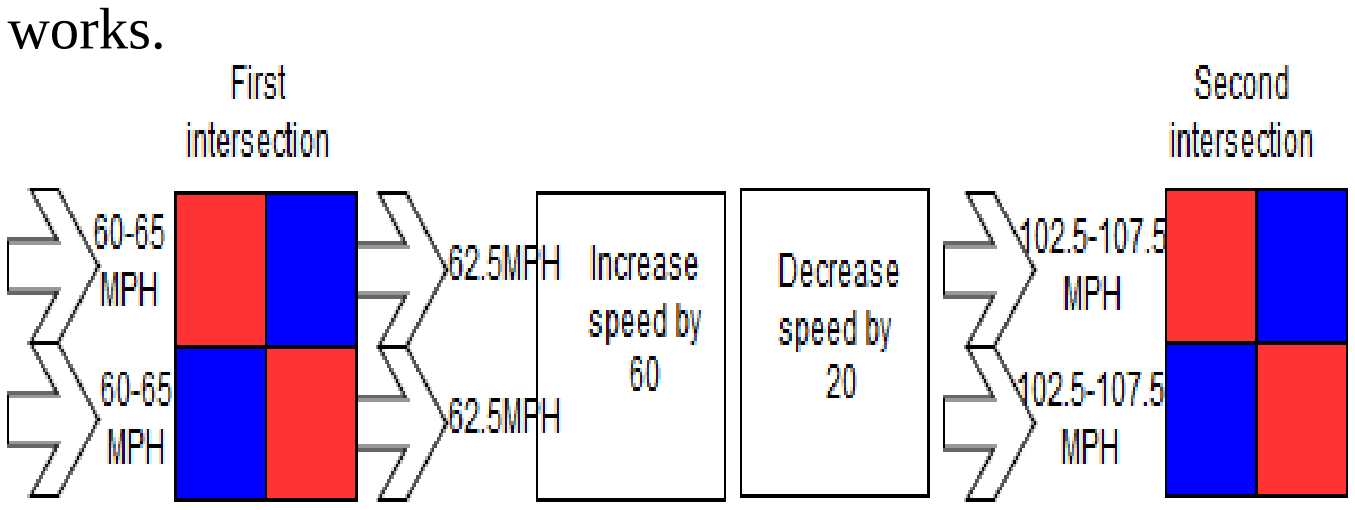

Figure11. Areas of changing the speed.

Regarding the third problem which is making a right turn. We made a classifier by using the key nearest neighbor (KNN) technique. In the KNN the intersection produce right turn containers based on whatever features we would like to apply. Some of the features that we could apply for the KNN: day, time, population of the city, event happening in the area,…etc. For our KNN we used the Euclidean distance function:

$$\sqrt{(q_1 - p_1)^2 + (q_2 - p_2)^2 + \cdots + (q_n - p_n)^2}$$

This classifier will be used to calculate the possibility of making a right turn for the next container (vehicle). Figure11 shows an example the KNN method.

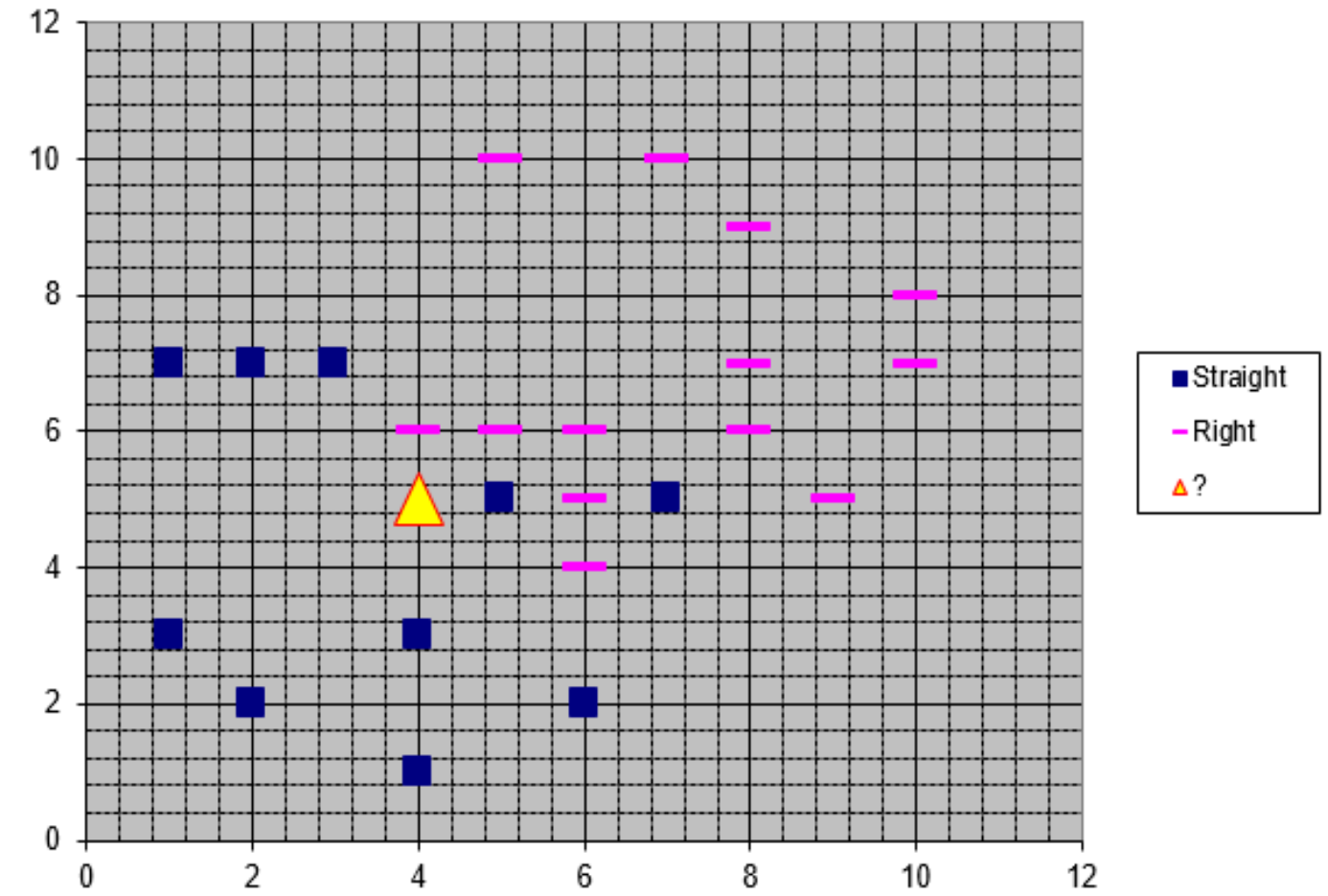

Figure11. The vehicle in the yellow triangle is going to make a right turn when K=3.

## IV. Simulation and Result

In our experiment we made a virtual intersection with the following features:

1- *Vehicles are coming from 4 different directions:*
   *A1: from east to west.*
   *A2: from west to east.*
   *B1: from north to west.*
   *B2: from west to north.*
2- *Vehicle arriving speed to the intersection is ranged between (60-65) mph.*
3- *The length for every container is 26.2467 ft.*
4- *The running time for the intersection is 1 minute.*
5- *The system produce one container for every opening lane.*
6- *Gates A1 and A2 should open together while gates B1 and B2 closed.*
7- *Gates B1 and B2 should open together while gates A1 and A2 closed.*
8- *Every lane has 60 containers in total.*

Figure12 below shows an illustration of the virtual intersection that we built for the experiment.

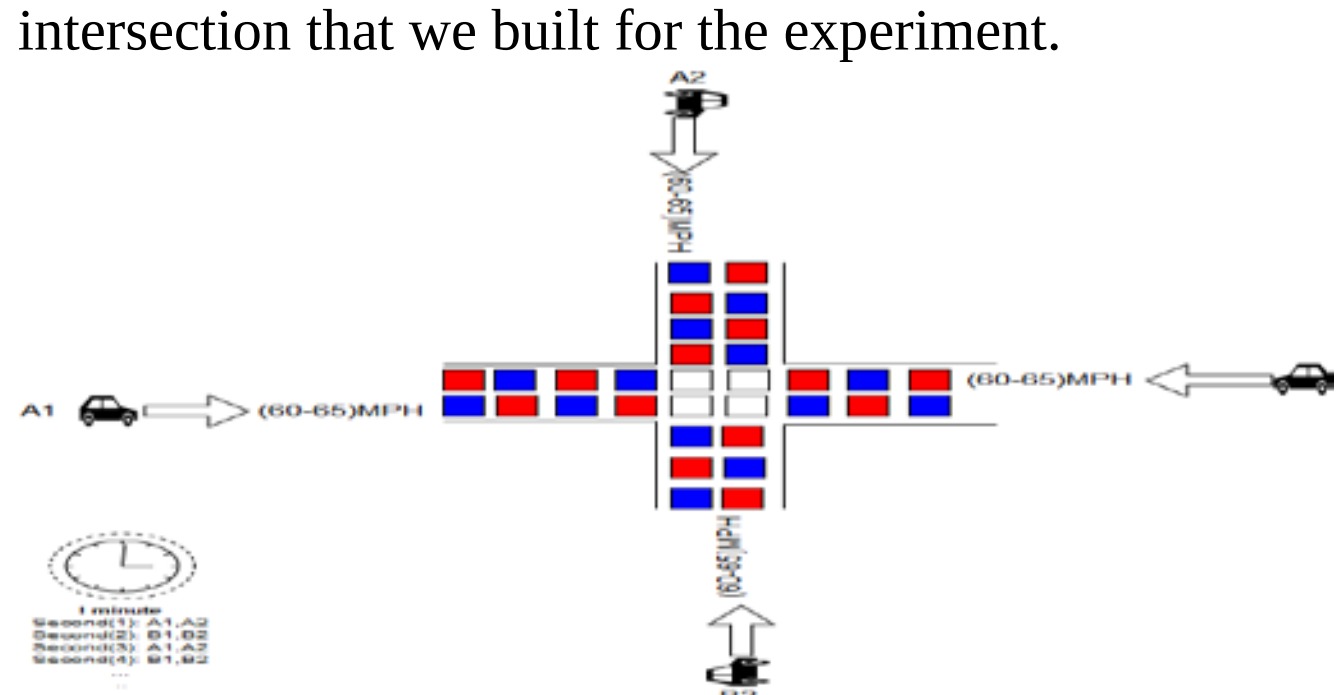

Figure12. Experiment virtual intersection.

For expecting the right turn in the KNN we need to enter initial testing data in the system. The following are the features and initial data:

| Day | Hour | Event in the area | Class of turn |
|---|---|---|---|
| 1 | 9 | 0 | + |
| 3 | 10 | 0 | + |
| 4 | 8 | 0 | + |
| 3 | 8 | 0 | + |
| 4 | 10 | 0 | + |
| 2 | 20 | 1 | - |
| 5 | 19 | 1 | - |
| 1 | 4 | 1 | - |
| 2 | 7 | 1 | - |

Where:

1- *Days are numbered from 1:Monday to 5:Friday.*
2- *Hour is a 24 format.*
3- *Event in the area is :*
   *0; if there is no event in the area of the intersection.*
   *1; if there is no event..*
4. *"+" means the vehicle is going to make a right turn. "-" means the vehicle is going in a straight direction.*

We also created 3 different patterns of the vehicles flow. The goal of these three patters is to calculate the waiting time if we got unordered flow. In addition we wanted to calculate the required space for each pattern. The first pattern represents the best case where the flow match the opening and the closing times of the intersection. The second pattern represents the worst case of the flow where vehicles keep coming even if the intersection is closed. The third and final pattern represent the random (normal) case where the flow is unpredictable.
Each of the three patterns has the follow features:

I. It has 60 spots.
II. Accept only one request per spot.
III. Runs for 1 minute.

For the patterns, we got different results based on the flow toward the intersection. For all of the following results:

*1 represents a request*
*0 represent no request*

In the average case where the flow pattern is: [1, 0, 1, 0, 1, 0, 1, 0, 1, 0, 1, 0,…..]: the arriving times for vehicles in one minute was: [0, 2, 4, 6, 8, 10, 12, 14, 16, 18, 20, 22, 24, 26, 28, 30, 32, 34, 36, 38, 40, 42, 44, 46, 48, 50, 52, 54, 56, 58]. We don't need to arrange them because they are already compatible with the opening and closing times for our lane. The waiting time in this case is always zero.

In the second case the flow is constant (worst case): [1, 1, 1, 1, 1, 1, 1, 1, 1, 1……]. The arriving times for this pattern in one minute was: [0, 1, 2, 3, 4, 5, 6, 7, 8, 9, 10, 11, 12, 13, 14, 15, 16, 17, 18, 19, 20, 21, 22, 23, 24, 25, 26, 27, 28, 29, 30, 31, 32, 33, 34, 35, 36, 37, 38, 39, 40, 41, 42, 43, 44, 45, 46, 47, 48, 49, 50, 51, 52, 53, 54, 55, 56, 57, 58, 59]. Arranging them to match the intersection opening and closing moments was a time consuming since it took 164.84 seconds as an average waiting time for each vehicle. Here is the order of arrivals after arranging them: [0, 2, 4, 6, 8, 10, 12, 14, 16, 18, 20, 22, 24, 26, 28, 30, 32, 34, 36, 38, 40, 42, 44, 46, 48, 50, 52, 54, 56, 58, 60, 62, 64, 66, 68, 70, 72, 74, 76, 78, 80, 82, 84, 86, 88, 90, 92, 94, 96, 98, 100, 102, 104, 106, 108, 110, 112, 114, 116, 118]. The worst part of this case is not only the high waiting time, but also we needed to increase the space by 100% to arrange the vehicles. The last pattern where the flow is unpredictable (random), the system was really promising since the waiting time was less than the old models. In our model, the waiting time was 35 seconds compared to 101 seconds for the old model. For testing, we created a random function that creates random requests. The requests we got were : [1, 1, 0, 1, 1, 1, 0, 0, 0, 1, 0, 1, 0, 1, 0, 1, 0, 1, 1, 0, 0, 1, 0, 0, 1, 0, 1, 0, 1, 0, 1, 0, 1, 1, 1, 0, 1, 0, 1, 0, 0, 1, 0, 1, 0, 0, 0, 0, 1, 0, 1, 0, 1, 1, 0, 0, 0, 1, 1, 1], and their arriving times were: [0, 1, 3, 4, 5, 9, 11, 13, 15, 17, 18, 21, 24, 26, 28, 30, 32, 33, 34, 36, 38, 41, 43, 48, 50, 52, 53, 57, 58, 59]. In this random pattern it only took 6.79 seconds as an average waiting time. Here is the arranged arrivals : [0, 0, 1, 0, 3, 0, 4, 0, 5, 0, 9, 0, 11, 0, 13, 0, 15, 0, 17, 0, 18, 0, 21, 0, 24, 0, 26, 0, 28, 0, 30, 0, 32, 0, 33, 0, 34, 0, 36, 0, 38, 0, 41, 0, 43, 0, 48, 0, 50, 0, 52, 0, 53, 0, 57, 0, 58, 0, 59]. To sum up the patterns result: -The first pattern (matched pattern) has 0 waiting time and 0 additional space. The second pattern (worst pattern) has increased by double in both timing and spacing. The third pattern (random flow) result depends on the number of requests: 0 if the number of requests below the giving spots, otherwise it will be ((n-30) /30*100) where n is the number of requests. However, it was less than the previous model in all tested random flow.

For predicting the right turn, whenever a new vehicle comes to the lane, its (day, hour, event) will be calculated by the KNN classifier to predict if it's going to make a right turn or not. The resulted data of the entered vehicle will be added to the classifier so it helps to predict the next generated spot. After running the simulation for one minute we got the following results:
Lane A1 and A2 instances:

| Day | Hour | Event in the area |
|---|---|---|
| 3 | 3 | 0 |
| 1 | 5 | 0 |
| 2 | 11 | 0 |
| 5 | 10 | 0 |
| 4 | 1 | 0 |
| 1 | 23 | 0 |
| 2 | 13 | 0 |
| 2 | 18 | 1 |
| 3 | 14 | 1 |
| 2 | 8 | 1 |
| 4 | 6 | 1 |
| 1 | 21 | 1 |
| 2 | 3 | 0 |
| 4 | 15 | 0 |
| 4 | 22 | 1 |
| 4 | 22 | 0 |
| 3 | 8 | 1 |
| 2 | 21 | 0 |
| 3 | 0 | 1 |
| 1 | 20 | 0 |
| 4 | 12 | 0 |
| 1 | 3 | 1 |
| 1 | 7 | 1 |
| 1 | 23 | 1 |
| 2 | 2 | 1 |
| 4 | 6 | 0 |
| 2 | 0 | 1 |
| 3 | 16 | 0 |
| 2 | 1 | 1 |
| 4 | 11 | 0 |

B1 and B2 instances:

| Day | Hour | Event in the area |
|---|---|---|
| 1 | 8 | 0 |
| 4 | 12 | 0 |
| 5 | 2 | 1 |
| 4 | 3 | 1 |
| 4 | 12 | 0 |
| 2 | 1 | 0 |
| 2 | 13 | 0 |
| 4 | 20 | 1 |
| 3 | 18 | 1 |
| 5 | 16 | 0 |
| 2 | 1 | 0 |
| 5 | 3 | 0 |
| 5 | 9 | 0 |
| 3 | 11 | 1 |
| 4 | 2 | 1 |
| 2 | 9 | 0 |
| 2 | 9 | 0 |
| 5 | 19 | 1 |
| 3 | 11 | 1 |
| 3 | 18 | 1 |
| 2 | 16 | 1 |
| 5 | 15 | 1 |
| 3 | 11 | 1 |
| 4 | 17 | 1 |
| 3 | 16 | 1 |
| 3 | 22 | 1 |
| 5 | 23 | 0 |
| 2 | 12 | 1 |
| 2 | 11 | 0 |
| 5 | 19 | 1 |

Turns prediction for lane A1 and lane A2:

| | | | | | | | | | |
|---|---|---|---|---|---|---|---|---|---|
| **1** | **2** | **3** | **4** | **5** | **6** | **7** | **8** | **9** | **10** |
| - | - | + | + | - | - | + | - | + | + |
| **11** | **12** | **13** | **14** | **15** | **16** | **17** | **18** | **19** | **20** |
| + | - | - | + | - | - | + | - | - | - |
| **21** | **22** | **23** | **24** | **25** | **26** | **27** | **28** | **29** | **30** |
| + | - | + | - | - | + | - | + | - | + |

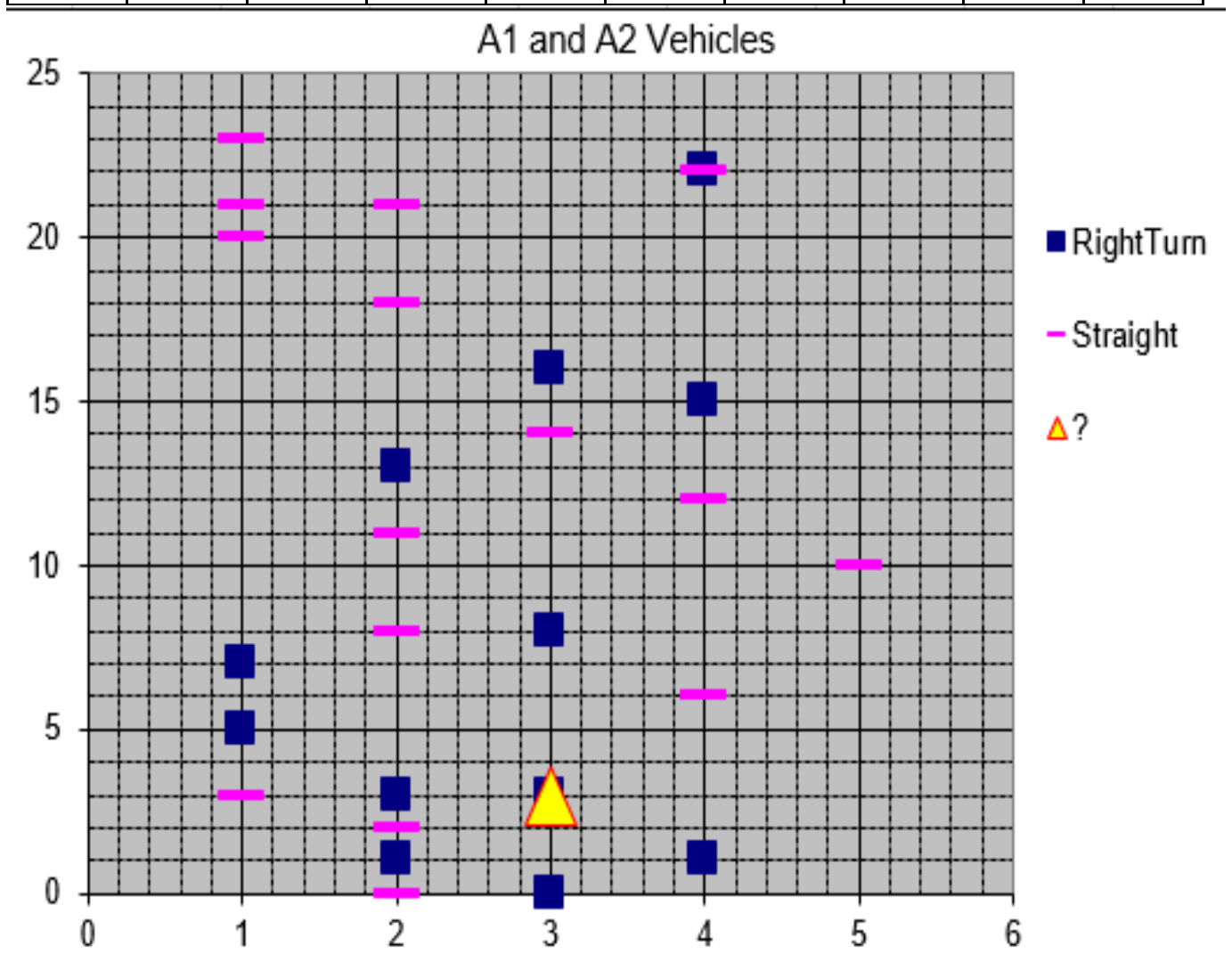

Figure14. KNN for lane A1 and A2.

Turns prediction for lane B1 and lane B2:

| | | | | | | | | | |
|---|---|---|---|---|---|---|---|---|---|
| **1** | **2** | **3** | **4** | **5** | **6** | **7** | **8** | **9** | **10** |
| + | + | - | - | + | - | + | - | - | - |
| **11** | **12** | **13** | **14** | **15** | **16** | **17** | **18** | **19** | **20** |
| - | - | + | + | - | + | + | - | + | - |
| **21** | **22** | **23** | **24** | **25** | **26** | **27** | **28** | **29** | **30** |
| - | - | + | - | - | - | - | + | + | - |

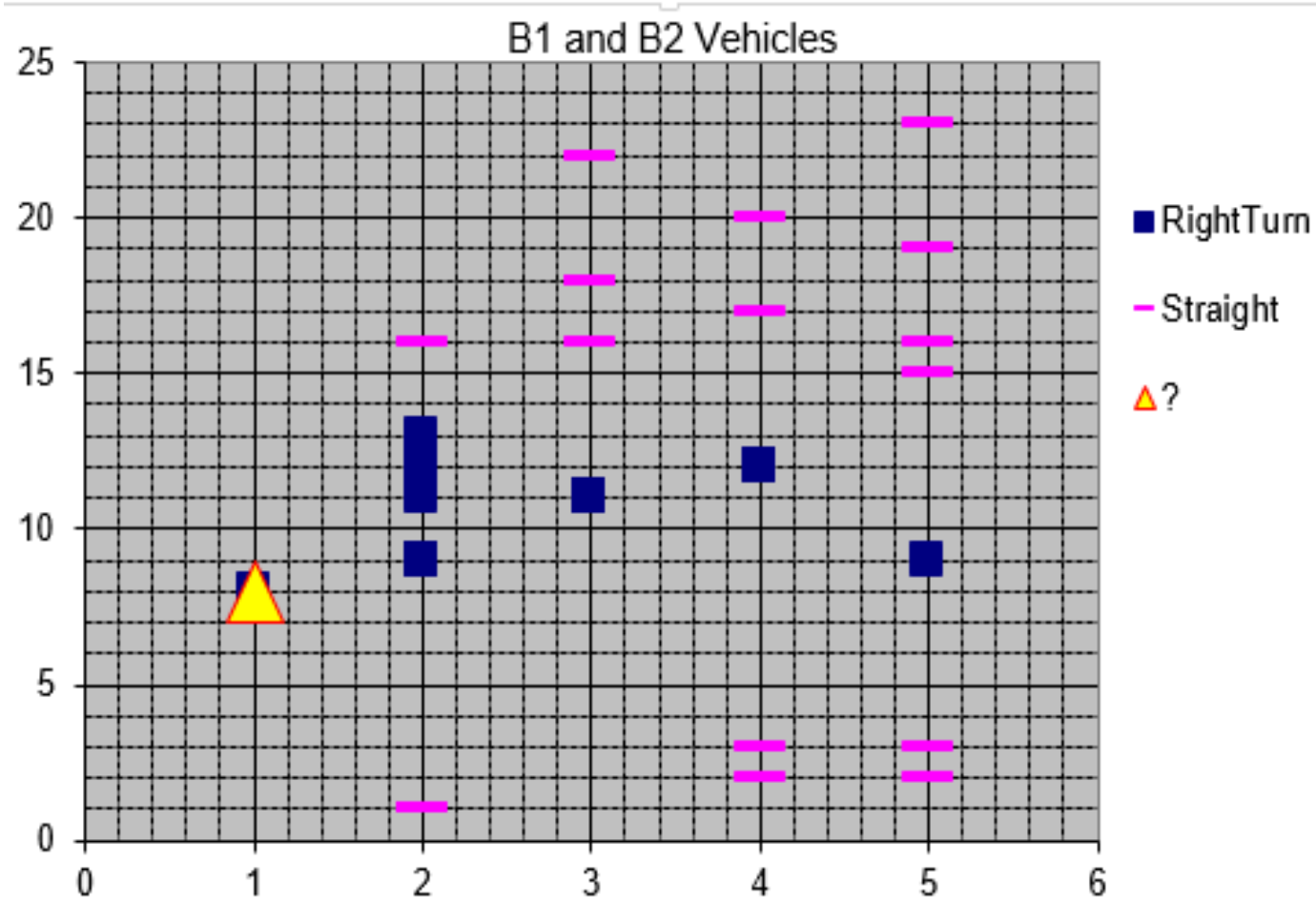

Figure14. KNN for lane B1 and B2.

The following Table1 shows the first 5 vehicles for each lane.

| Vehicle ID | Lane | Arriving Time | Right Turn | Exiting Time |
|---|---|---|---|---|
| 123 | A1 | 0.0 | No | 17.179657557103365 |
| 206 | A2 | 0.0 | No | 17.179657557103365 |
| 301 | B1 | 1.0 | Yes | 18.179657557103365 |
| 413 | B2 | 1.0 | Yes | 18.179657557103365 |
| 106 | A1 | 2.0 | No | 19.179657557103365 |
| 224 | A2 | 2.0 | No | 19.179657557103365 |
| 306 | B1 | 3.0 | Yes | 20.179657557103365 |
| 405 | B2 | 3.0 | Yes | 20.179657557103365 |
| 115 | A1 | 4.0 | Yes | 21.179657557103365 |
| 200 | A2 | 4.0 | Yes | 21.179657557103365 |
| 324 | B1 | 5.0 | No | 22.179657557103365 |
| 422 | B2 | 5.0 | No | 22.179657557103365 |
| 109 | A1 | 6.0 | Yes | 23.179657557103365 |
| 213 | A2 | 6.0 | Yes | 23.179657557103365 |
| 311 | B1 | 7.0 | No | 24.179657557103365 |
| 418 | B2 | 7.0 | No | 24.179657557103365 |
| 105 | A1 | 8.0 | No | 25.179657557103365 |
| 222 | A2 | 8.0 | No | 25.179657557103365 |
| 313 | B1 | 9.0 | Yes | 26.179657557103365 |
| 418 | B2 | 9.0 | Yes | 26.179657557103365 |

Table1. Data of the first 5 vehicles that entered the intersection.

## V. Conclusion

The proposed autonomous scheduling system introduced in this paper is based on the manufacturing production line system. The intersection environment is sat up in advance before any vehicle arrives to the intersection. We tested three different patterns (best-worst-random). In the first pattern where the flow matches the intersection requirements, the queue waiting time was zero. In the worst case the waiting time was high and could be presented as X!. In the third pattern the waiting time was promising for a real environment since it got less waiting time compared to the old model. Having said that, using a huge amount of space is still an issue that we are working on for the second version of the system.